\documentclass{article}
\usepackage{authblk} 
\usepackage{graphicx}
\usepackage[utf8]{inputenc}
\usepackage[T1]{fontenc}
\usepackage{bm}
\usepackage[a4paper, total={7in, 8in}]{geometry}
\usepackage{multicol}
\usepackage{blindtext}
\usepackage{cite}
\usepackage{xcolor}
\usepackage{amsmath}
\usepackage{amsfonts} 
\usepackage{mathtools}
\usepackage{nccmath}
\usepackage{siunitx}

\title{Surface Phonon Polariton Ellipsometry}
\author[1]{Giulia Carini}
\author[1]{Richarda Niemann}
\author[1]{Niclas Sven Mueller}
\author[1]{Martin Wolf}
\author[1]{Alexander Paarmann}

\affil[1]{Fritz Haber Institute of the Max Planck Society, Berlin, Germany}

\begin{document}

\flushbottom
\maketitle

\begin{abstract}
     Surface phonon polaritons (SPhPs) have become a key ingredient for infrared nanophotonics, owing to their long lifetimes and the large number of polar dielectric crystals supporting them. While these evanescent modes have been thoroughly characterized by near-field mapping or far-field intensity measurements over the last decade, far-field optical experiments also providing phase information are less common. In this paper, we study surface phonon polaritons at the gallium phosphide (GaP)-air interface in the momentum domain using the Otto-type prism coupling geometry. We combine this method with spectroscopic ellipsometry to obtain both amplitude and phase information of the reflected waves across the entire reststrahlen band of GaP. By adjusting the prism-sample air gap width, we systematically study the dependence of the ellipsometry parameters on the optical coupling efficiency. In particular, we show that the combined observation of both ellipsometry parameters - amplitude and phase - provides a powerful tool for the detection of SPhPs, even in the presence of high optical losses. Finally, we theoretically study how surface phonon polariton ellipsometry can reveal the emergence of vibrational strong coupling through changes in the topology of their complex plane trajectories, opening up a new perspective on light-matter coupling.
    
\end{abstract}

\begin{multicols}{2}           
        
    \section{Introduction}
  
        Phonon polaritons (PhPs) are light-matter electromagnetic waves that arise in polar crystals from the hybridization of infrared (IR) photons with IR-active phonons. The research field has grown out of efforts to explore the possibilities of nanophotonics beyond the limits imposed by plasmonics\cite{maier2007plasmonics} by extending the available spectral range for polaritonic modes into the IR and THz, while dramatically reducing optical losses thanks to the long lifetimes of phonons (typically on the picosecond timescale)\cite{caldwell2015low,caldwell2013low, khurgin2018relative}. Furthermore, PhPs have attracted much attention in recent years due to the high directionality resulting from the inherently anisotropic optical response of low-symmetry crystals, which are natural platforms for these types of modes\cite{passler2022hyperbolic,ni2023observation}. In addition, the wide variety of crystals available in nature allows to explore the extreme multifacetedness of PhPs\cite{galiffi2024extreme}, paving the way for the development of a plethora of applications for advanced nanophotonic-based devices\cite{gubbin2022surface}.
    
        The coupling between photons and phonons is responsible for the formation of bulk PhPs and leads to an avoided crossing in their dispersion relation. This, in turn, results in the formation of a spectral region known as reststrahlen band (RB), bounded by the transversal optical (TO) and longitudinal optical (LO) phonon frequencies, where light propagation within the crystal bulk is prohibited\cite{falge1973dispersion, huber2005near, neuner2009critically}. This frequency interval is therefore characterized by a near-unity reflectance, reminiscent of the optical response of metals below the plasma frequency\cite{caldwell2015low}. 
        In polar crystals, the Rabi splitting $\Omega_\text{R}$ for bulk phonon polaritons - related to the RB width through $\Omega_R = \sqrt{\omega_{\text{LO}}^2-\omega_{\text{TO}}^2}$ - is generally in the order of $10$ to $80\,\%$ of the TO frequency\cite{mueller2020deep, barra2021microcavity}, which naturally places PhPs in the ultrastrong light-matter coupling regime. 
        
        In recent years, much effort was dedicated to modulating light-matter coupling, e.g., with different types of photonic resonators\cite{frisk2019ultrastrong}, with crystals of plasmonic nanoparticles \cite{mueller2020deep} or with molecular switches embedded in optical microcavities\cite{thomas2020new}.
        However, achieving such modulation employing PhPs in polar crystals is not so easy from an experimental point of view. It requires manipulating the LO-TO splitting by irreversibly changing the material properties, e.g., by isotopic substitution\cite{giles2018ultralow}, or alternatively employing heterostructured metamaterials\cite{matson2024role,ratchford2019controlling,chen2023van}. 
        

        Interestingly, due to their "metal-like" optical behaviour in the RB, polar dielectrics can support evanescent surface waves known as surface phonon polaritons (SPhPs) in that spectral region. The evanescent nature of SPhPs exhibiting momenta outside the light cone prevents their direct optical excitation from free space. However, it is possible to excite SPhPs within the RB employing dedicated experimental schemes to overcome the momentum mismatch with free-space radiation\cite{passler2017second,passler2018strong,ratchford2019controlling,runnerstrom2018polaritonic,folland2020vibrational,kohlmann2022second,ni2023observation}. Recent studies have investigated different surface modes in polar dielectric crystals and heterostructures with intensity measurements enabled by grating\cite{kohlmann2022second,folland2020vibrational} or prism\cite{passler2018strong, ratchford2019controlling, runnerstrom2018polaritonic,ni2023observation} coupling techniques. In particular, when arranged in the so-called Otto geometry\cite{otto1968excitation}, prism coupling provides unsurpassed control over the excitation efficiency of the SPhPs\cite{passler2017second,neuner2009critically}. Therefore, Otto-type prism coupling offers a unique platform for the investigation of light-matter coupling in polar crystals.         

        In this work, we combine Otto-type prism coupling  with spectroscopic ellipsometry, a widespread technique that is typically employed to derive the complex dielectric tensors of crystals\cite{schubert2016anisotropy}.
        This combined approach enables the acquisition of both amplitude and phase information of SPhPs here exemplified in the far-infrared for gallium phosphide (GaP). While the optical response of this crystal is already well known\cite{streyer2014engineering}, to the best of our knowledge, no phase-resolved optical response of its surface modes has been reported.
        By means of this hybrid method, which we call  'surface phonon polariton ellipsometry', we measure the complex-valued ellipsometry parameter $\rho$ for various  optical coupling efficiencies, as controlled in the experiment by varying the air gap width between the prism and the sample. The results show that the representation of the quantity $\rho$ in the complex plane provides a unique perspective on optical coupling phenomena, e.g., by allowing a complementary visualization of the critical coupling condition\cite{neuner2009critically,passler2017second}. We compare these data with transfer matrix simulations of free-space excited GaP, where no evanescent waves emerge, providing a complementary insight into light-matter coupling both in free space and mediated by evanescent waves. Finally, we theoretically apply our method to study phonon-polariton strong coupling using thin films of MoS$_{\text{2}}$ on GaP as a potential vibrational strong coupling platform, and predict a change in topology of the complex plane trajectories of $\rho$ by varying the MoS$_{\text{2}}$ thickness, conceptually similar to previous ellipsometry work in the visible\cite{thomas2020new}. From a broader perspective, the experimental access to the optical phase of surface modes opens up new ways of thinking about strong light-matter coupling, the emergence of SPhPs within the RB, and the coupling of these surface modes to other resonances.     
        
       
    \section{Experimental results}
\begin{figure}[ht!]                     \includegraphics[width=1.0\textwidth]{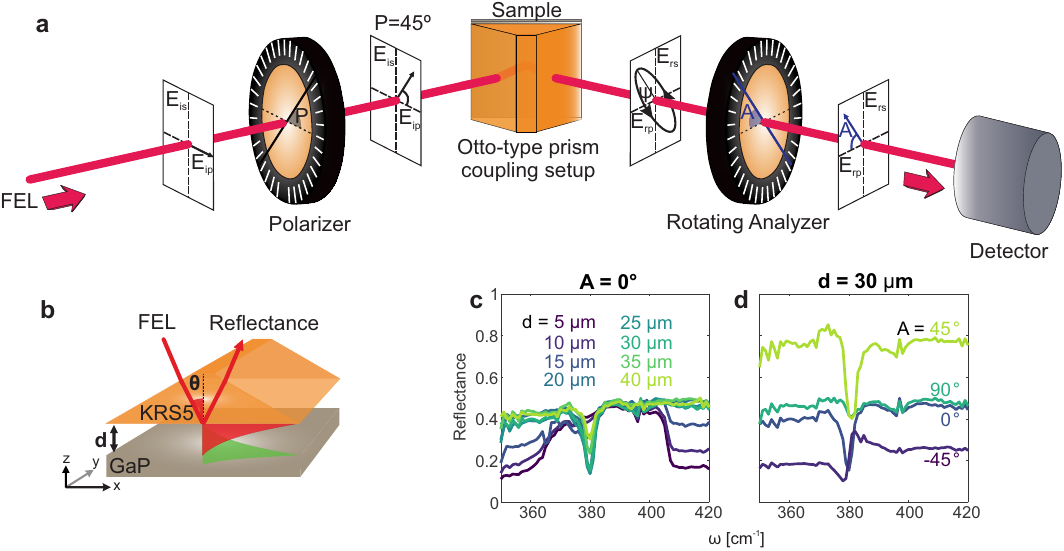}
    \caption{\textbf{Experimental setup for spectroscopic ellipsometry measurements in an Otto-type prism coupling configuration}. (a) Schematic of the rotating-analyzer ellipsometry (PSA$_\text{R}$) setup employed in the experiment, consisting of two wire grid polarizers and the prism-sample system. The red- and green-shaded areas denote the evanescent waves for TIR and SPhP, respectively. (b) Otto-type prism coupling geometry employed for the excitation and detection of surface phonon polaritons at the interface between gallium phosphide and air. (c) Experimental far-infrared reflectance spectra recorded at eight different air gap widths for a fixed analyzer angle \textit{A}\,=\,0$^\circ$. (d) Experimental reflectance spectra recorded at four different analyzer angle configurations for a fixed air gap width \textit{d}\,=\,30\,\textmu m. 
    }
    
    \label{fig:fig1}
\end{figure}

        We applied far-IR spectroscopic ellipsometry for phase-resolved investigations of SPhP in bulk GaP. In order to overcome the momentum mismatch between the incident light and the SPhP for excitation of the surface modes, we employed a high-refractive index Thallium Bromo-Iodide (KRS5) prism arranged in the Otto-type prism coupling geometry \cite{falge1973dispersion, otto1968excitation, neuner2009critically,passler2017second}. This technique allows to deterministically tune the evanescent in-plane momentum of the excitation by changing the incidence angle of the laser beam, as well as the coupling efficiency by adjusting the spacing $d$ between the prism and the sample. This air gap width is precisely monitored  by means of white-light interferometry\cite{pufahl2018controlling}. When far-IR radiation is incident on the prism-air interface at an angle $\theta$ larger than the critical condition for total internal reflection (TIR), the excitation of SPhPs in the underlying substrate can occur. This is schematically illustrated in Fig.\,\ref{fig:fig1}b by the overlap between the red- and the green-shaded areas, which denote the evanescent waves for total internal reflection and SPhP, respectively. In our experiment, we perform narrow-band (< 0.5\,\% bandwidth) excitation spectroscopy in the far-IR region of the electromagnetic spectrum (350-420\,cm$^{\text{-1}}$) by utilizing the  IR free-electron laser at the Fritz-Haber Institute (FHI-FEL)\cite{schollkopf2015new} as tunable IR light source.

        \end{multicols}

        \begin{multicols}{2}

        In a typical reflectance measurement at the GaP surface in the Otto geometry, the SPhP resonance results in a dip in the reflectance spectrum with the dip magnitude representing the SPhP amplitude\cite{streyer2014engineering,neuner2009critically}. To obtain the phase information, however, we employed a home-built rotating-analyzer ellipsometry (also called polarizer-sample-rotating analyzer, PSA$_{\text{R}}$) setup. As depicted in Fig.\,\ref{fig:fig1}a, the setup consists of two wire-grid polarizers in addition to the prism-sample system. The incident p-polarized far-IR radiation is converted by a linear polarizer - which rotates the polarization by an angle $P$ of $45^\circ $ - into a linear combination of p- and s-polarization components with zero phase offset. Then, the light encounters the Otto geometry, where a non-zero relative phase $\Delta$ between the two components arises due to the different Fresnel reflection coefficients for the two polarization components at the interface between the GaP substrate and air, in general leading to an elliptical polarization of the reflected beam. To fully characterize the polarization of the reflected beam, reflectance spectra at four analyzer settings $A\, = \,0^\circ,\,90^\circ,\,45^\circ,\,-45^\circ$ are recorded with a pyroelectric detector, for various air gap widths at a fixed incidence angle $\theta$ of $27^\circ$. The complete set of these reflectance spectra is available in the Supplementary Information (Figs.\,S1\,a-d).
        
        In Fig.\,\ref{fig:fig1}\,c, we show the experimental data recorded at $A=0^\circ$ which illustrates the p-polarized component of the elliptically polarized light reflected off the Otto system. The data is referenced to unity reflectance at total reflection with very large ($\approx 100\, \mu$m) air gap width, and renormalized to the maximum reflectance obtained from simulations (further details on the data treatment in the SI, Section 1). In the overcoupled regime, i.e., at small prism-sample distances, we can clearly recognize the highly reflective RB region of GaP. In this frequency range, where light propagation inside the crystal is prohibited, the evanescent p-polarized waves at the prism-sample gap couple to SPhPs, resulting in pronounced dips in the reflectance spectra, as visible in the green or turquoise curves in Fig.\,\ref{fig:fig1}\,c (\textit{d}\,=\,25,\,30\,\textmu m, respectively). Notably, Maxwell's equations do not predict s-polarized near-fields in the case of optically isotropic crystals such as GaP\cite{zayats2005nano}. Hence, when comparing the reflectance spectra at different analyzer settings for an air-sample spacing \textit{d}\,=\,30\,\textmu m in Fig.\,\ref{fig:fig1}\,d, we do not expect to observe any SPhP excitation in the turquoise curve ($A\, = \,90^\circ$), which corresponds to the spectrum of the s-polarized component, where indeed the SPhP resonance dip is largely suppressed in the experiment. The small remnant feature may arise from imperfect polarization states.
        
        All four reflectance measurements displayed in Fig.\,\ref{fig:fig1}\,d are required to experimentally obtain the four-component Stokes vector $\vec{S}$\cite{fujiwara2007spectroscopic}. In fact, only the first three Stokes vector components can be measured directly in our PSA$_{\text{R}}$ setup, as they are obtained from the quantities plotted in Fig.\,\ref{fig:fig1}\,d, as follows:
        \begin{equation}\begin{split}
            &S_{\text{0}} = R_{A=0^\circ}+R_{A=90^\circ}\text{,}\\
            &S_{\text{1}}  = R_{A=0^\circ}-R_{A=90^\circ}\text{,} \\
            &S_{\text{2}}  = R_{A=45^\circ}-R_{A=-45^\circ}\text{.}
            \end{split}
        \end{equation}      
        $S_{\text{3}}$ - which defines the latitude in the Poincar\'e sphere representation and contains information about the degree of ellipticity and the chirality - is derived from the other components by assuming totally polarized light as $S_3 = \sqrt{S_0^2~-~S_1^2~-~S_2^2}$\cite{fujiwara2007spectroscopic}. From the Stokes vector, we extract the ellipsometry parameters $\Psi$ and $\Delta$, which correspond to amplitude  and phase, respectively, of the complex ratio $\rho = r_{\text{p}}/r_{\text{s}}$ where  $r_{\text{p}}$ and $r_{\text{s}}$ are the Fresnel reflection coefficients for p- and s-polarization, respectively
        \begin{equation}
        \label{rho}
            \rho ~ = ~ \frac{r_{\text{p}}}{r_{\text{s}}} ~ = ~ \tan(\Psi)~e^{i\Delta}\text{.}
        \end{equation}

		The quantities $\Psi$ and $\Delta$ are related to the Stokes parameters as follows\cite{fujiwara2007spectroscopic}:
		\begin{equation}
            \label{ellips_par_Psi}
            \begin{split}
				\Psi~ = ~-\frac{1}{2} ~\textrm{arccos}\left(-\frac{S_1}{S_0}\right)\text{,}\\
			\end{split}
		\end{equation} 

        \begin{equation}
            \label{ellips_par_Delta}
            \centering
            \Delta~ =~
            \left\{\begin{split}
                &\textrm{arctan}\left(-\frac{S_3}{S_2}\right) ~~~~~~ ~~\textrm{for} ~~~ \cos(\Delta) > 0\text{,} \\
                &\textrm{arctan}\left(-\frac{S_3}{S_2}\right)+\pi ~~~ \textrm{for} ~~~ \cos(\Delta) < 0, ~\sin(\Delta)\geq0\text{,} \\ 
                &\textrm{arctan}\left(-\frac{S_3}{S_2}\right)-\pi ~ ~~\textrm{for} ~~~ \cos(\Delta) < 0, ~\sin(\Delta)<0 \text{.} \\ 
            \end{split}\right.
         \end{equation}
        More details about spectroscopic ellipsometry in this particular configuration can be found in Ref. \cite{fujiwara2007spectroscopic}.

\begin{figure}[ht!] 
\centering\includegraphics[width=1\textwidth]{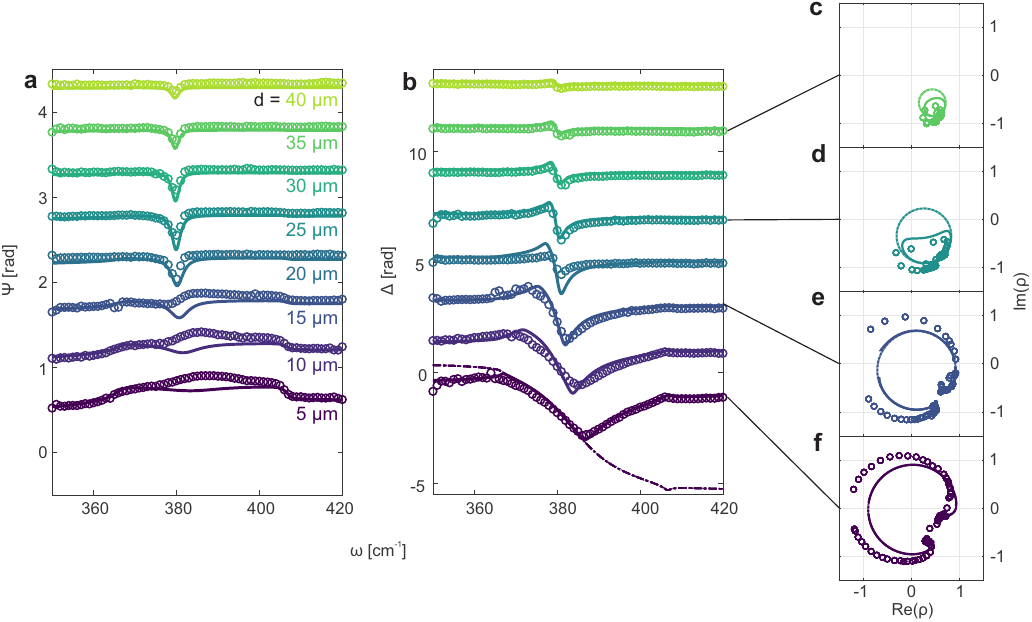}
    \caption{\textbf{Ellipsometry parameters $\Psi(\omega)$ and $\Delta(\omega)$ extracted from experimental data  (circles) and TM simulations (solid and dashed lines)}. (a,b) Spectra of the amplitude $\Psi$ (a) and phase difference $\Delta$ between s- and p-polarization components (b) for selected gap widths \textit{d}. When performing the TM simulations, the presence of a Thallium Bromoiodide (KRS5) prism was considered while setting up the material system. An offset was added both to the incidence angle and the gap width, resulting in the following choice of parameters for the simulated spectra: $\theta_{\text{sim}}$\,=26.55\,$^\circ$ and \textit{d}\textsubscript{sim}\,=\,10,\,15,\,20,..\textmu m. (c-f) Complex plane representation of $\rho$, plotted as a function of frequency. The simulations shown in the panels were obtained using two different methods. The dashed lines have been plotted directly from the ratio of the Fresnel reflection coefficients as described by Eq.\,(\ref{rho}). This also applies to the dashed purple line in panel b, which is mirrored for the purpose of display (the mirroring is responsible for the mismatch of the two curves at $\omega$\,<\,370\,cm$^{\text{-1}}$). On the other hand, the solid lines are the result of simulations applied to mimic the PSA$_{\text{R}}$ setup, taking additionally into account the broadening of the FEL frequency (as explained in detail in the SI).
}
    \label{fig:fig2}
\end{figure}

        The spectra of the ellipsometry parameters $\Psi$ and $\Delta$ extracted from the experimental data are shown as waterfall plots in Fig.\,\ref{fig:fig2}\,a,b (circles) for various air-gap widths $d$. In Fig.\,\ref{fig:fig2}\,c-f we show the corresponding representation of $\rho$ in the complex plane for selected values of \textit{d}. In general, plotting $\rho$ in the complex plane allows us to look at the SPhP excitation from an unusual perspective that is not possible with traditional intensity measurements. These data are overlaid with the corresponding transfer matrix (TM) simulations\cite{passler2017generalized} (solid lines). 
        As already extensively discussed in a previous work by Passler \textit{et al.}\cite{passler2017second}, the gap dependence of the polaritonic feature's prominence in the reflectance spectrum exhibits a critical behaviour that is reproduced here in the amplitude $\Psi$ data. In Fig.\,\ref{fig:fig2}\,a, we observe a faint dip at the smallest gap width of 5\,\textmu m, where the coupling efficiency is strongly diminished due to high radiative losses of the SPhP evanescent tail back into the prism. For larger gaps, an increasingly prominent and narrow dip emerges in the spectrum, until the condition of critical coupling is reached at \textit{d}\textsubscript{crit}\,$\approx$\,25\,\textmu m, corresponding to the optimal distance at which the incident light is completely absorbed by the SPhP. At a prism-sample spacing larger than the critical gap width (\textit{d}\,>\,\textit{d}\textsubscript{crit}), we access the 'undercoupled' regime, where the small overlap between the two evanescent waves shown in the schematics of Fig.\,\ref{fig:fig1}\,b prevents an efficient SPhP excitation. This results in a shallower, yet narrow dip in the spectrum of $\Psi$ (see, e.g., the light turquoise and green curves in panel Fig.\,\ref{fig:fig2}\,a, \textit{d}\,=\,35,\,40\,\textmu m). Due to irregularities in the shape of the baseline (see SI Section 1), the agreement between experiment and simulation is not optimal at small gaps, but improves significantly with increasing gap width, and is excellent at the critical coupling condition. 

        \end{multicols}

        \begin{multicols}{2}

        A similar critical behaviour is observed for the phase $\Delta$ (Fig.\,\ref{fig:fig2}\,b), as well as in the selected plots of $\rho$ in the complex plane (Fig.\,\ref{fig:fig2}\,c-f). The absorption of IR light results in dips in the amplitude spectra (Fig.\,\ref{fig:fig2}\,a), which translate into phase jumps in $\Delta$ (Fig.\,\ref{fig:fig2}\,b) and into quasi-circular trajectories in the complex plane representation of $\rho$ (Fig.\,\ref{fig:fig2}\,c-f). At small gaps in the overcoupled regime, the phase changes nearly linearly across the full RB over a total range of $2\pi$, corresponding to the full loop in Fig.\,\ref{fig:fig2}\,f. This behaviour is displayed by the dashed lines in Figs.\,\ref{fig:fig2}\,b-f (in panel b, only one 'unfolded' $\Delta$-spectrum has been reported, which has been mirrored for the purpose of display). Notably, the experimental data in this case is folded since the 
        PSA$_{\text{R}}$ method according to Eq.~\ref{ellips_par_Delta} cannot resolve the ambiguity in the sign of $S_{\text{3}}$, due to which $-\pi$\,<\,$\Delta$\,<\,0\cite{fujiwara2007spectroscopic} (see SI section 2 for an extensive discussion). Upon increasing the air gap, the dispersive region of $\Delta$ shrinks and the curve steepens to form a Lorentzian-like SPhP resonance, with the steepest slope emerging at critical coupling (see $\Delta_{\text{theory}}$ in Fig.\,S3\,c). The critical coupling condition is hence characterized by the highest gradient of the $\Delta$-spectra at the SPhP resonance frequency (see the turquoise-green curve at \textit{d}\,=\,25\,\textmu m, in panel b), which corresponds to the point where the loop formed by $\rho$ intersects with the origin in the complex plane. Notably, close to the critical coupling condition (Fig.\,\ref{fig:fig2}\,d), the finite spectral line width of our FEL results in a distortion of the signals leading to a deformation of the loop (see Fig.\,S4). For even larger gaps, the magnitude of the resonance diminishes, associated with a further shrinking of the loop size in the complex plane. In this case, the trajectories are defined only in half of the complex plane (when $\Im(\rho)$\,<\,0) such that $\Delta$ no longer exhibits a $2\pi$ step across the RB.

        Notably, the critical behaviour visible in the gap width dependence of $\Psi$ and $\Delta$, when considered independently in panels a and b, transforms into a monotonic change of the loop radius in the complex plane. A very similar behavior of RB-induced loops of $\rho$ in the complex plane can be found when analyzing free-space infrared spectroscopic ellipsometry data of polar crystals such as ZnO\cite{ashkenov2003infrared} or alumina\cite{schubert2000infrared}. However, in free-space ellipsometry the loops only cover a half-space of the complex plane corresponding to a $\pi$-step in $\Delta$ only. In order to better understand these complex plane loops conceptually, we will analyze the different scenarios theoretically in the following.



    \section{Complex plane loops in polariton ellipsometry}

\begin{figure}[ht!] 
\centering\includegraphics[width=1\textwidth]{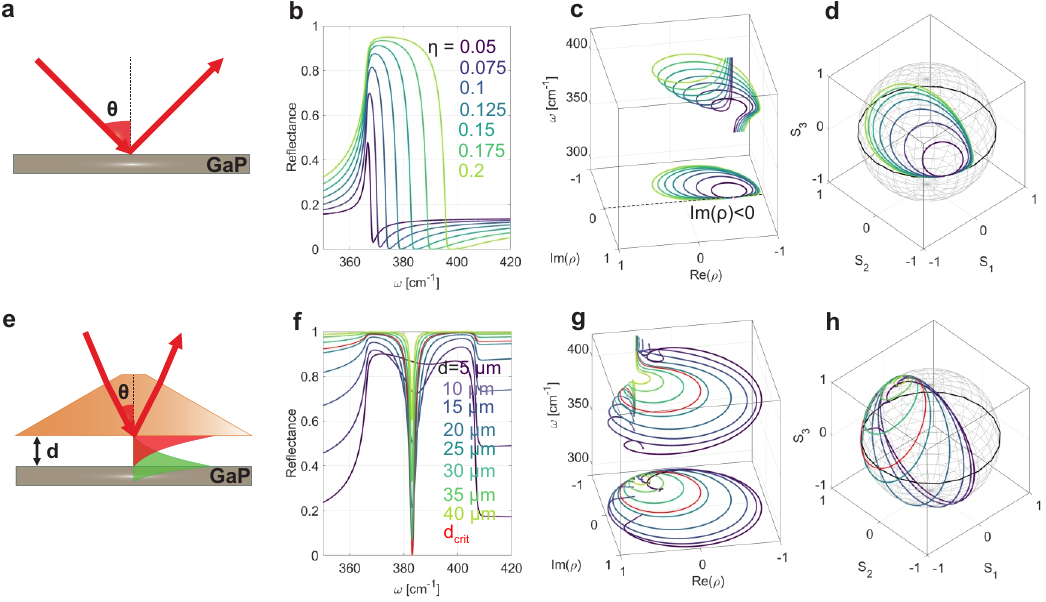}
    \caption{\textbf{TM simulations of ellipsometry parameter for the GaP-air interface in free-space (upper row) and in a total internal reflection-based illumination scheme (lower row).} (a,e) Sketches illustrating the reflection scheme in free-space (a) and in the prism coupling geometry (e). (b,f) Simulated reflectance spectra around the reststrahlen band (RB) of GaP. The simulations are performed considering incidence angles $\theta$ of $45^\circ$ (b) and $27^\circ$ (f). The various curves shown in the panels correspond to different coupling strength \textit{g} (b) and gap widths \textit{d} (f). (c,g) 3D-plots of $\rho$ vs $\omega$ and corresponding projections in the \textit{xy}-plane of loops in the complex plane representation. (d,h) Poincar\'e-sphere representation of the polarization state of the light reflected off the air-GaP interface, without (d) and with (h) the prism. The red curves in panels (f-h) correspond to the critical coupling condition, and were calculated by assuming a prism-sample spacing of 26.53\,\textmu m. This value was obtained numerically from the Parratt recursion formula for a multilayer system after setting the condition for critical coupling: $\Im(\rho)=\Re(\rho)=0$. The black line at the equator in panels (d) and (h) is marked for reference.   
    }
    \label{fig:fig3}
\end{figure}

        To gain a deeper understanding of the loops of $\rho$ in the complex plane, we now theoretically compare the Otto-type prism coupling configuration with free-space excitation of a polar crystal using TM calculations, see Fig.\,\ref{fig:fig3}. For the free-space case, we assume that light is incident on the air-sample interface at an angle of $45^\circ$ (see Fig.\,\ref{fig:fig3}\,a). 
        In this picture, a strong material resonance like a TO phonon couples to the incoming light, which is then responsible for the excitation of the upper or the lower bulk phonon polariton (BPhP) branch. The dielectric permittivity of the substrate can be described by means of a classical Lorentz model with an oscillator strength $g$:
        \begin{equation}
            \varepsilon(\omega) = \varepsilon_\infty+ g\frac{\omega_{\text{TO}}^2}{\omega_{\text{TO}}^2 - \omega^2-i\Gamma\omega}   
        \end{equation}
        By modifying $g$, it is possible to tune the Rabi splitting $2\Omega_\text{R}$ of the system, since $g=\frac{4 \epsilon_{\infty} \Omega_\text{R}^2}{\omega_{\text{TO}}^2}$, and thus the normalized coupling strength $\eta=\frac{\Omega_\text{R}}{\omega_{\text{TO}}}$ \cite{mueller2020deep}. Fig.\,\ref{fig:fig3}\,b shows the reflectance spectra calculated for various values of $\eta$, from strong to ultra-strong ($\eta>0.1$) light-matter coupling regimes\cite{frisk2019ultrastrong}. More details on the simulations are available in SI Section 3.  In these plots, the high-reflectance spectral region becomes broader at increasing normalized coupling strengths $\eta$. Notably, for all the curves there is a frequency range where the real part of the dielectric permittivity acquires negative values, leading to the buildup of a RB in all the cases analyzed here. If we study the frequency dependence of $\rho$ and visualize it in a 3D plot, as shown in Fig.\,\ref{fig:fig3}\,c, we observe a helicoidal shape that results in loops when projected onto the complex $\rho$ plane. The loop size increases with $\eta$. For this system, $\rho$ can undergo a maximum phase jump of $\pi$, which causes the asymmetric trajectories to be confined to the negative half of the plane (where $\Im(\rho)<0$). This behavior is typical of a dielectric environment in the presence of a resonance\cite{jurna2010visualizing}. 
        
        An alternative way of presenting the loops is to plot the polarization state of the light reflected off the air-GaP interface on the Poincar\'e sphere (see Fig.\,\ref{fig:fig3}\,d). Under the assumption of totally polarized light, the Stokes vector traces squeezed loops in the northern hemisphere. This visualization method helps to reveal the physical meaning of the trajectories, in particular showing that the strong coupling of the light to a dipolar material resonance is not sufficient to induce a change in the handedness of the reflected radiation (see SI Section 4 for a comparison between the two visualization methods). The fact that $S_3$ is always positive determines that the light is right-elliptically polarized over the entire frequency range considered, and for all values of $\eta$.

        The picture described above does not include SPhPs. To do so, we now add a high refractive-index prism as sketched in Fig.\,\ref{fig:fig3}\,e in the TM simulations. In this scenario, the surface mode emerges as a dip in the reflectance spectra in Fig.\,\ref{fig:fig3}\,f. Like in the experiments shown in Fig.\,\ref{fig:fig2}, we tune the air gap width between the prism and the sample to modulate the coupling efficiency. The red curve corresponds to the critical coupling condition, where every photon impinging on the surface is absorbed at the SPhP resonance. 

        \end{multicols}

        \begin{multicols}{2}
        Furthermore, the complex ratio $\rho$ is plotted in 3D as a function of frequency (analogous  to Fig.\,\ref{fig:fig3}\,c), shown in Fig.\,\ref{fig:fig3}\,g for various gap widths. The projections of $\rho$ in the complex plane are quasi-circular trajectories which undergo a transition from the undercoupled to the overcoupled regime, crossing the origin of the complex plane at the critical coupling condition (red curve). This is in stark contrast to the loops discussed in the panel above (Fig.\,\ref{fig:fig3}\,c) obtained from the free-space reflection scheme, which solely emerge in the negative half-space of the complex plane. As observed experimentally, when a KRS5 prism is brought close enough to the sample in the overcoupled regime, the phase $\Delta$ covers the full range of $2\pi$ (see Fig.\,S3\,c) which causes the loops across the entire complex plane. The experimental results shown in Figs.\,\ref{fig:fig2}\,c-f - despite some experimental limitations - confirm this transition. 
        
        Interestingly, the data shows that $\rho$ crosses the origin of the complex plane even in the presence of high radiative losses. Consider for instance the amplitude $\Psi$ measured at \textit{d}\,=\,5\,\textmu m (Fig.\,\ref{fig:fig2}\,a), where the dip is broad and hardly recognizable. The corresponding loop plotted in Figs.\,\ref{fig:fig2}\,f, however, has a large radius and extends over the entire complex plane. As a complementary picture, Fig.\,\ref{fig:fig3}\,h provides an overview of the polarization state of the reflected radiation. Notably, the Stokes vector traces circular trajectories on the Poincar\'e sphere that cross the equator at small gap widths. This indicates that an efficient surface mode's excitation - independently of the radiative losses - is responsible for a change of handedness in the polarization state of the reflected beam.

        While manipulating the LO-TO splitting ($\propto \eta$) in a bulk crystal is not experimentally feasible without structurally modifying the material properties, the Otto-type prism coupling technique offers a unique non-invasive tunability of the optical coupling efficiency. We show that, in combination with spectroscopic ellipsometry, this approach enables an advanced control over the SPhP excitation, opening new pathways for the study of the light-matter coupling and the emergence of surface modes in the photonic bandgap.
        
    \section{Vibrational Strong Coupling}
\begin{figure}[ht!] 
\centering\includegraphics[width=1\textwidth]{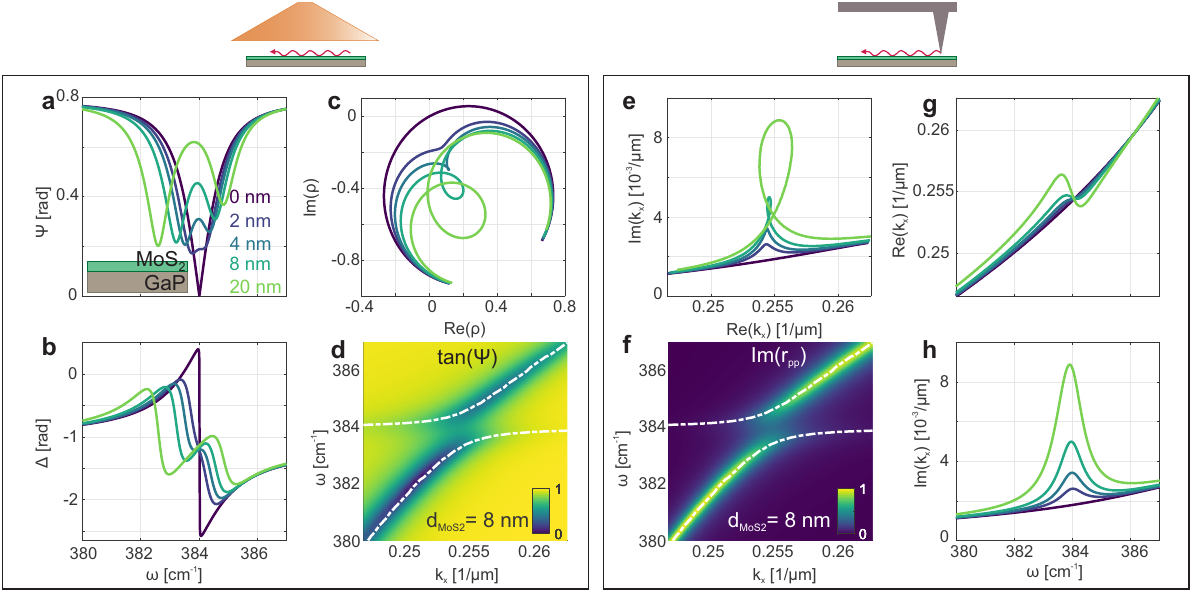}
    \caption{\textbf{Transition from weak to strong coupling in a MoS$_{\text{2}}$/GaP system for different MoS$_{\text{2}}$ thicknesses}. The large black boxes group the panels according to the two different methods for studying strong coupling: TIR and tip-based approaches on the left and right, respectively, as shown in the sketches above. (a,b,c) Simulated ellipsometry parameters for a material system of the type depicted in the inset of panel a. The different curves are obtained by varying the thickness of the MoS$_{\text{2}}$ thin slab, see labels. The simulations were performed by assuming an incidence angle $\theta$\,=\,27.165$^\circ$ and a gap \textit{d}\,=\, 24.95\,\textmu m. (d,f) Simulated reflectivity maps derived from the magnitude of $\rho$ ($\tan(\Psi)$) and the imaginary part of the reflection coefficient for p-polarized light ($\Im(r_{\text{pp}})$) normalized to its maximum, respectively. Both maps are obtained by assuming a MoS$_{\text{2}}$ thickness of 8 nm. The simulations are overlaid with semi-analytical dispersion curves (dashed white lines); see SI Section 6. (e,g,h) Real (g) and imaginary (h) parts of the in-plane wave vector $k_\text{x}$ obtained from the peak positions and line widths, respectively, of horizontal line cuts through the dispersion maps (exemplified in panel f for \textit{d}\,=\,8\,nm). Whilst in panels g and h $\Re({k_{\text{x}}})$ and $\Im({k_{\text{x}}})$ are plotted against $\omega$, panel e shows a plot of the imaginary part against the real part. 
    }
    \label{fig:fig4}
\end{figure}

        The unique perspective on optical coupling phenomena arising in the complex plane of the ellipsometry parameter $\rho$ may also provide opportunities for vibrational strong coupling for infrared sensing applications\cite{autore2018boron,folland2020vibrational,dolado2022remote}. To explore these phenomena using polariton ellipsometry, we theoretically consider in the following a simple model system comprising of a very thin film of the two-dimensional van-der Waals crystal MoS$_{\text{2}}$ placed on top of a GaP substrate. MoS$_{\text{2}}$ exhibits a weak IR-active phonon resonance at $\omega_{\text{TO}} = 384$\,cm$^{-1}$ within the SPhP band of GaP\cite{wieting1971infrared}. Thus, we can expect strong coupling of the GaP SPhP to the MoS$_{\text{2}}$ phonon that can be characterized in the Otto configuration. A schematic of the material stack is shown in the inset of Fig.\,\ref{fig:fig4}\,a.
        
        TM simulations were performed by varying the thickness of the MoS$_{\text{2}}$ layer from 0 to 20 nm at a fixed angle of incidence and at the critical coupling condition for the GaP SPhP without MoS$_{\text{2}}$, see Fig.\,S6 for details. The angle of incidence was chosen such that the SPhP resonance matched exactly to the MoS$_{\text{2}}$ phonon ($\theta\,=\,27.165^{\circ}$). The resulting ellipsometry parameters $\Psi$ and $\Delta$ (derived from $\rho=r_{\text{pp}}/r_{\text{ss}}$) are plotted in Fig.\,\ref{fig:fig4}\,a,b. Although the dip in $\Psi$ (the phase jump in $\Delta$) shows a broadening and a deformation (a flatter slope and a break in the monotonicity) already for a 2 nm thick MoS$_{\text{2}}$ slab, the onset of strong coupling can only be observed at a thickness of 4 nm (see SI Section 4 for a more detailed explanation). As the MoS$_{\text{2}}$ slab thickness increases, two dips in the amplitude $\Psi$ appear and become increasingly distinct in Fig.\,\ref{fig:fig4}\,a. For the phase $\Delta$ in Fig.\,\ref{fig:fig4}\,b, the interaction between the two modes manifests itself in the emergence of two Lorentzian-like resonances that become gradually more spectrally separated (see e.g. turquoise and green lines in Fig.\,\ref{fig:fig4}\,a,b).
        The amplitude and phase information, when considered independently, do not provide any robust criterion to characterize the strong coupling regime without first quantifying the Rabi splitting and comparing it with the line widths of the uncoupled resonances.
        
        
        However, the combined observation of both ellipsometry parameters ($\Psi$ and $\Delta$) in the complex plane visualization of $\rho$ provides a more suitable tool to detect the onset of strong coupling\cite{thomas2020new}, see Fig.\,\ref{fig:fig4}\,c. In these plots, the transition from weak to strong coupling with increasing thickness is manifested as a change in topology. At \textit{d}\textsubscript{MoS$_{\text{2}}$}\,=\,2\,nm, the area enclosed by the primary loop - which, as discussed in detail in the previous section, represents the signature of the surface phonon polariton of GaP - reduces, while its continuity is interrupted by a kink. At the onset of strong coupling the dimple transforms into a secondary loop right around \textit{d}\textsubscript{MoS$_{\text{2}}$}\,=\,4\,nm, which becomes more pronounced for thicker MoS$_{\text{2}}$ slabs (\textit{d}\textsubscript{MoS$_{\text{2}}$}\,=\,8,\,20\,nm). A similar topological transition in the optical response has been recently observed with spectroscopic ellipsometry in a coupled system consisting of optical microcavities and organic molecules\cite{thomas2020new}. 
        

        To confirm that our model system has indeed already approached at \textit{d}\textsubscript{MoS$_{\text{2}}$}\,=\,8\,nm the strong coupling regime, we plot the dispersion in Fig.\,\ref{fig:fig4}\,d. There we show the TM-simulated color map of tan($\Psi$). The strong coupling with the associated avoided crossing in the dispersion relation is reproduced by the overlaid semi-analytical model (white dashed lines), see SI section 6 for details. The upper and lower polariton branches are the eigenvalues of the Hopfield hamiltonian, derived by assuming a coupling strength \textit{g} larger than the system's losses. Thus, surface phonon polariton ellipsometry promises to emerge as a powerful tool to study vibrational strong coupling, since it enables the observation of the transition into strong coupling as change in the topology of $\rho$, providing a complementary platform for vibrational strong coupling employing SPhPs\cite{autore2018boron,folland2020vibrational}.

        \end{multicols}

        \begin{multicols}{2}

        Interestingly, a similar approach of exploring strong coupling through loops in the complex plane was recently discussed for the complex-valued in-plane polariton wave vector $k_{\text{x}}$ \cite{kusch2021strong,bylinkin2021real}. In this case, simultaneous near-field measurements of the polariton wavelength and propagation length also enable a complex plane representation, in which strong coupling results in the buildup of a loop\cite{kusch2021strong}. For a direct comparison to our ellipsometry approach, we extracted the respective quantities from TM simulations of the same system, see Fig.\,\ref{fig:fig4}\,e-h. Trajectories of the complex-valued wave vector are shown in Fig.\,\ref{fig:fig4}\,e. The real and imaginary parts of $k_{\text{x}}$ are extracted from the peak positions and line widths, respectively, of horizontal cuts through dispersion maps of $\Im(r_{\text{pp}})$, as exemplified in Fig.\,\ref{fig:fig4}\,e for \textit{d}\textsubscript{MoS$_{\text{2}}$}\,=\,8\,nm. The results for $\Re({k_{\text{x}}})$ and $\Im({k_{\text{x}}})$ for the same MoS$_{\text{2}}$ layer thicknesses are shown in Figs.\,\ref{fig:fig4}\,g,h. Notably, previous works\cite{kusch2021strong, bylinkin2021real, torma2014strong} have already discussed the possibility that the presence of a strongly coupled resonance (the MoS$_{\text{2}}$ TO phonon, in our case) in the spectral region of the propagating polariton introduces a back-bending in the real part of $k_x$. The back-bending can hence be interpreted as a signature of strong coupling \cite{wolff2018modal}(Fig.\,\ref{fig:fig4}\,g). This feature starts becoming prominent at \textit{d}\textsubscript{MoS$_{\text{2}}$}\,=\,8\,nm and corresponds to an increase in oscillator strength, accompanied by a peak in the propagation losses, as can be seen in Fig.\,\ref{fig:fig4}\,h. Furthermore, when the in-plane wave vector is plotted in the complex plane in Fig.\,\ref{fig:fig4}\,e, the curves take on a looping shape at the strong coupling transition, reminiscent of the loops observed using the ellipsometry method (see Fig.\,\ref{fig:fig4}\,d).

        Both methods analyzed in this section are suitable for the study of coupled systems and provide a qualitative criterion for strong coupling between independent material resonances (see SI section 5 for the study of the two approaches in the context of light-matter coupling). The near-field approach has been experimentally demonstrated\cite{kusch2021strong, bylinkin2021real} using interferometric near-field imaging techniques capable of resolving the complex-valued polariton momentum separately for each real frequency. On the other hand, polariton spectroscopy techniques, such as the prism coupling, typically probe the complex polariton frequency through their finite line width for a given real wave vector, and the dispersion they can map does not feature backbending\cite{passler2018strong}. This makes these techniques incompatible with and complementary to the complex momentum analysis. The representation of the parameter $\rho$ provided by spectroscopic ellipsometry, as introduced here, opens new avenues for detecting vibrational strong coupling also with far-field spectroscopic techniques.

    \section{Conclusions}
        In summary, in this work we demonstrated a new method for retrieving amplitude and phase information for surface phonon polaritons by combining the Otto-type prism coupling technique with spectroscopic ellipsometry. By varying the air gap width between the prism and the GaP sample, we systematically investigated the behavior of the ellipsometry parameters upon varying the optical coupling efficiency. Analysing the complex quantity $\rho$ as extracted from the ellipsometry parameters provides new insight into the critical coupling behaviour, showing a continuously increasing loop radius when moving from the undercoupled to the overcoupled regime, crossing the origin of the complex plane at critical coupling. To gain a deeper understanding of these circular trajectories in the complex plane, we theoretically compared the evanescent excitation via the prism coupling geometry with a free-space excitation with varying oscillator strength of the underlying phonon resonance. Finally, we theoretically studied the feasibility of using surface phonon polariton ellipsometry to characterize vibrational strong coupling. We observed the transition from weak to strong coupling for MoS$_{\text{2}}$ ultra-thin films on GaP through the emergence of a secondary loop in $\rho$ with increasing thickness of the MoS$_{\text{2}}$. Therefore, polariton ellipsometry and in particular complex plane visualization of $\rho$  presents a compelling opportunity by providing a qualitative, topology-based criterion for identifying vibrational strong coupling. More generally, this work demonstrates surface phonon polariton ellipsometry as a powerful tool for the characterization of SPhPs and the analysis of light-matter coupling of materials.  

    \section{Acknowledgements}
        We thank M. Schubert and A. Ruder (U Nebraska) for helpful discussions in the initial phase of this project. We also thank Joanna Urban and Olga Minakowa (FHI Berlin) for fruitful discussions that contributed to the development of this work. Finally, we acknowledge W. Sch\"{o}llkopf and S. Gewinner (FHI-FEL) for proving a stable FEL beam thoughout the whole measurement time necessary for this experiment.


\bibliographystyle{naturemag}
\bibliography{bibliography}
\end{multicols}

\end{document}